\documentclass[aps,prd,groupedaddress,amssymb,showpacs,showkeys]{revtex4}
\begin{document}
\def\bib#1{[{\ref{#1}}]}
\title{\bf Space-time deformations as extended conformal transformations}

\author{S. Capozziello$^{a,b}$ and  C. Stornaiolo$^b$ }
\email{capozziello@na.infn.it, stornaiolo@na.infn.it}

\affiliation{$^a$ Dipartimento di Scienze Fisiche, Universit\`a di
Napoli "Federico II", and $^b$INFN, Sez. di Napoli, Compl. Univ.
di Monte S. Angelo, Edificio G, Via Cinthia, I-80126 - Napoli,
Italy }
\date{\today}

\begin{abstract}
A definition of space-time metric deformations on an
$n$-dimensional manifold is given. We show that such deformations
can be regarded as extended conformal transformations. In
particular,  their features can be related to the perturbation
theory giving a natural picture by which gravitational waves are
described by small deformations of the metric. As further result,
deformations can be related to approximate Killing vectors
(approximate symmetries) by which it is possible to parameterize
the deformed region of a given manifold. The perspectives and some
possible physical applications of such an approach are discussed.
\end{abstract}



 \maketitle

\section{Introduction}
The issue to consider a general way to deform the space-time
metrics is not new. It has been posed in different ways and is
related to several physical problems ranging from the spontaneous
symmetry breaking of unification theories up to gravitational
waves, considered as space-time perturbations. In cosmology, for
example, one faces the problem to describe an observationally
lumpy universe at small scales which becomes isotropic and
homogeneous at very large scales according to the Cosmological
Principle. In this context, it is crucial to find a way to connect
background and locally perturbed metrics \cite{Ellis84}. For
example, McVittie \cite{mcvittie} considered a metric which
behaves as a Schwarzschild one at short ranges and as a
Friedman-Lemaitre-Robertson-Walker metric at very large scales.
Gautreau \cite{gautreau} calculated the metric generated by  a
Schwarzschild mass embedded in a Friedman cosmological fluid
trying to address the same problem. On the other hand, the
post-newtonian parameterization, as a standard, can be considered
as a deformation of a background, asymptotically flat Minkowski
metric.

In general,  the deformation problem has been explicitly  posed by
Coll and collaborators \cite{Coll1,Coll:2001wy,Llosa:2003di} who
conjectured the possibility to obtain any metric from the
deformation of a space-time with constant curvature. The problem
was solved only for  3-dimensional spaces but a straightforward
extension should be to achieve the same result for space-times of
any dimension.

In principle, new exact solutions of the Einstein field equations
can be obtained by studying perturbations. In particular, dealing
with perturbations as Lorentz matrices of scalar fields
$\Phi^{A}_{\phantom{A}C}$ reveals particularly useful. Firstly
they transform as scalars with respect the coordinate
transformations. Secondly, they are dimensionless and, in each
point, the matrix $\Phi^{A}_{\phantom{A}C}$ behaves as the element
of a group. As we shall see below, such an approach can be related
to the conformal transformations giving an "extended"
interpretation  and a straightforward physical meaning of them
(see \cite{faraoni,allemandi} and references therein for a
comprehensive review). Furthermore scalar fields related to
space-time deformations have a straightforward physical
interpretation which could contribute to explain several
fundamental issues  as the Higgs mechanism in unification
theories, the inflation in cosmology and  other  pictures where
scalar fields play a fundamental role in dynamics.

In this paper, we are going to discuss  the properties of the
deforming matrices $\Phi^{A}_{\phantom{A}C}$ and we will derive,
from the Einstein equations, the field equations for them, showing
how them can parameterize the  deformed metrics, according to the
boundary and initial conditions and to the energy-momentum tensor.

The layout of the paper is the following. In Sec.II, we define the
space-time perturbations in the framework of the metric formalism
giving the notion of first and second deformation matrices.
Sec.III is devoted to the main properties of deformations. In
particular, we discuss how deformation matrices can be split in
their trace, traceless and skew parts. We derive the contributions
of deformation to the geodesic equation and, starting from the
curvature Riemann tensor, the general equation of deformations. In
Sec.IV we discuss the notion of linear perturbations under the
standard of deformations. In particular, we recast the equation of
gravitational waves and the transverse traceless gauge under the
standard of deformations. Sec.V is devoted to discuss the action
of deformations on the Killing vectors. The result consists in
achieving a notion of approximate symmetry. Discussion and
conclusions are given in Sec.VI. In  Appendix, we discuss in
details how deformations act on affine connections.

\section{Generalities on space-time deformations}

In order to start our considerations, let us take into account a
metric $\mathbf{g}$ on a space-time manifold $\mathcal{M}$. Such a
metric is assumed to be an exact solution of the Einstein field
equations. We can decompose it by a co-tetrad field $\omega^{A}(x)
$
\begin{equation}\label{tetrad}
\mathbf{g}=\eta_{AB}\omega^{A} \omega^{B}.
\end{equation}
Let us define now a new tetrad field
$\widetilde{\omega}=\Phi^{A}_{\phantom{A}C}(x)\,\omega^{C}$, with
$\Phi^{A}_{\phantom{A}C}(x)$ a matrix of scalar fields. Finally we
introduce a space-time  $\widetilde{\mathcal{M}}$ with the metric
$\widetilde{g}$  defined in the following way
\begin{equation}\label{deformed}
\mathbf{\widetilde{g}}=\eta_{AB}\Phi^{A}_{\phantom{A}C}\Phi^{B}_{\phantom{B}D}\,\omega^{C}
\omega^{D} = \gamma_{CD}(x)\omega^{C}
\omega^{D},
\end{equation}
where also $\gamma_{CD}(x)$ is a matrix of fields which are
scalars with respect to the coordinate transformations.

If $\Phi^{A}_{\phantom{A}C}(x)$ is  a Lorentz matrix in any point of  $\mathcal{M}$, then
\begin{equation}\label{i.3}
    \widetilde{g}\equiv g
\end{equation}
otherwise  we say that $\widetilde{g}$ is a  deformation of $g$
and $\widetilde{\mathcal{M}}$ is a deformed $\mathcal{M}$. If all
the functions of $\Phi^{A}_{\phantom{A}C}(x)$ are continuous, then
there is a {\it one - to - one} correspondence between the points
of $\mathcal{M}$ and the points of $\widetilde{\mathcal{M}}$.

In particular, if  $\xi$ is a Killing vector for $g$ on
$\mathcal{M}$, the  corresponding vector $\widetilde{\xi}$ on
$\widetilde{\mathcal{M}}$ could not  necessarily be a Killing
vector.

A particular subset of these deformation matrices is given by
\begin{equation}\label{conformal}
{\Phi}{^{A}_{C}}(x)=\Omega(x)\, \delta^{A}_{\phantom{A}C}.
\end{equation}
which define conformal transformations of the metric,
\begin{equation}\label{confmetric}
  \widetilde{g}=\Omega^{2}(x) g\,.
\end{equation}

In this sense, the  deformations defined by Eq. (\ref{deformed})
can be regarded as a generalization of the conformal
transformations.

We  call the matrices $\Phi^{A}_{\phantom{A}C}(x)$ {\it first
deformation matrices}, while we can refer to
\begin{equation}\label{seconddeformation}
 \gamma_{CD}(x)=\eta_{AB}\Phi^{A}_{\phantom{A}C}(x)\Phi^{B}_{\phantom{B}D}(x).
\end{equation}
as the {\it second deformation matrices}, which, as seen above,
are also matrices of scalar fields. They generalize  the Minkowski
matrix \(\eta_{AB}\) with constant elements in the definition of
the metric. A further restriction on the matrices
$\Phi^{A}_{\phantom{A}C}$ comes from the theorem proved by Riemann
by which an $n$-dimensional metric has $n(n-1)/2$ degrees of
freedom (see \cite{Coll:2001wy} for details). With this
definitions in mind, let us consider the main properties of
deforming matrices.

\section{Properties of  deforming matrices}

Let us take into account a four dimensional space-time with
Lorentzian signature. A family of matrices
$\Phi^{A}_{\phantom{A}C}(x)$ such that
\begin{equation}\label{fi}
    \Phi^{A}_{\phantom{A}C}(x)\in GL(4)\, \forall x,
\end{equation}
are defined on such a space-time.

These functions are not necessarily continuous and can connect
space-times with different topologies. A singular scalar field
introduces a deformed manifold $\widetilde{\mathcal{M}}$ with a
space-time singularity.

As it is well known, the Lorentz matrices
$\Lambda^{A}_{\phantom{A}C}$ leave the Minkowski metric invariant
and then

\begin{equation}\label{deformlambda2}
\mathbf{g}=\eta_{EF}\Lambda^{E}_{\phantom{E}A}\Lambda^{F}_{\phantom{F}B}\Phi^{A}_{\phantom{A}C}\Phi^{B}_{\phantom{B}D}\,
\omega^{C}
\omega^{D} =\eta_{AB}\Phi^{A}_{\phantom{A}C}\Phi^{B}_{\phantom{B}D}\,\omega^{C}
\omega^{D}.
\end{equation}
It follows that $\Phi^{A}_{\phantom{A}C}$ give rise to right
cosets of the Lorentz group, i.e. they are the elements of  the
quotient group $GL(4,\mathbf{R})/SO(3,1)$. On the other hand,  a
right-multiplication of $\Phi^{A}_{\phantom{A}C}$ by a Lorentz
matrix induces a different deformation matrix.

The inverse deformed metric is
\begin{equation}\label{inversemetric}
     \widetilde{g}^{ab}=\eta^{CD}{\Phi^{-1}}^{A}_{\phantom{A}C}{\Phi^{-1}}^{B}_{\phantom{B}D}e_{A}^{a}e_{B}^{b}
\end{equation}
where ${\Phi^{-1}}^{A}_{\phantom{A}C}{\Phi}^{C}_{\phantom{C}B}=\Phi^{A}_{\phantom{A}C}{\Phi^{-1}}^{C}_{\phantom{C}B}=
\delta^{A}_{B}$.

Let us decompose now the matrix $\Phi_{AB}=\eta_{AC}\,
\Phi^{C}_{\phantom{C}B}$ in its symmetric and antisymmetric parts
\begin{equation}\label{decomposition}
     \Phi_{AB}= \Phi_{(AB)}+\Phi_{[AB]}= \Omega\,\eta_{AB} +  \Theta_{AB} + \varphi_{AB}
\end{equation}
where $ \Omega= \Phi^{A}_{\phantom{A}A}$,  $ \Theta_{AB} $ is the
traceless symmetric part   and $ \varphi_{AB}$ is the skew
symmetric part of  the first deformation matrix respectively. Then
standard conformal transformations are nothing else but
deformations with $\Theta_{AB}=\varphi_{AB}=0$ \cite{wald}.

Finding the inverse matrix ${\Phi^{-1}}^{A}_{\phantom{A}C}$ in
terms of $\Omega$, $\Theta_{AB}$  and $\varphi_{AB}$ is not
immediate, but as above, it can be split in the three terms
\begin{equation}\label{inversesplitting}
      {\Phi^{-1}}^{A}_{\phantom{A}C}=\alpha\delta^{A}_{\phantom{A}C}+\Psi^{A}_{\phantom{A}C}+\Sigma^{A}_{\phantom{A}C}
\end{equation}
where $\alpha$, $\Psi^{A}_{\phantom{A}C}$ and
$\Sigma^{A}_{\phantom{A}C}$ are  respectively the trace, the
traceless symmetric part and the antisymmetric part of the inverse
deformation matrix. The second deformation matrix, from the above
decomposition, takes the form
\begin{equation}\label{secondmatrix}
  \gamma_{AB}= \eta_{CD}(\Omega\, \delta_{A}^{C}+
  \Theta_{\phantom{C}A}^{C}+ \varphi_{\phantom{C}A}^{C})(\Omega\, \delta_{B}^{D}+
  \Theta_{\phantom{D}B}^{D}+ \varphi_{\phantom{D}B}^{D})
\end{equation}
and then

$$   \gamma_{AB}=  \Omega^{2}\,\eta_{AB} + 2\Omega\,\Theta_{AB}+  \eta_{CD}\, \Theta_{\phantom{C}A}^{C}\,\Theta_{\phantom{D}B}^{D} + \eta_{CD}\, (\Theta_{\phantom{C}A}^{C}\,\varphi_{\phantom{D}B}^{D}$$
\begin{equation}\label{secondmatrix1}
  +   \varphi_{\phantom{C}A}^{C}\,\Theta_{\phantom{D}B}^{D}) + \eta_{CD}\,\varphi_{\phantom{C}A}^{C}\,\varphi_{\phantom{D}B}^{D}.
\end{equation}
In general, the deformed metric can be split as
\begin{equation}\label{splits}
     {\tilde{g}}{_{ab}}=\Omega^{2}{g}{_{ab}}+{\gamma}{_{ab}}
\end{equation}
where
$${\gamma}{_{ab}}=\left( 2\Omega\,\Theta_{AB}+  \eta_{CD}\, \Theta_{\phantom{C}A}^{C}\,\Theta_{\phantom{D}B}^{D} + \eta_{CD}\, (\Theta_{\phantom{C}A}^{C}\,\varphi_{\phantom{D}B}^{D}+   \varphi_{\phantom{C}A}^{C}\,\Theta_{\phantom{D}B}^{D})\right. $$
\begin{equation}\label{acca}
  \left. + \eta_{CD}\,\varphi_{\phantom{C}A}^{C}\,\varphi_{\phantom{D}B}^{D}\right){\omega}{^{A}_{a}}{\omega}{^{B}_{b}}
  \end{equation}
In particular,  if $ \Theta_{AB}=0$, the deformed metric
simplifies to
\begin{equation}\label{split}
    \widetilde{g}_{ab}=\Omega^{2}g_{ab}+\eta_{CD}\,\varphi_{\phantom{C}A}^{\,C}\,\varphi_{\phantom{D}B}^{\,D}
    \omega^{A}_{\phantom{A}a}\omega^{B}_{\phantom{B}b}
\end{equation}
and, if $\Omega=1$, the deformation of a metric consists in adding
to the background metric a tensor $\gamma_{ab}$. We have to
remember that all these quantities are not independent as, by the
theorem mentioned in \cite{Coll:2001wy}, they have to form at most
six independent functions in a four dimensional space-time.

Similarly the controvariant deformed metric can be always
decomposed in the following way
\begin{equation}\label{controvariantdecomposition}
    \widetilde{g}^{ab}= \alpha^{2}g^{ab}+ \lambda^{ab}
\end{equation}
Let us find the relation between ${\gamma}{_{ab}}$ and $
\lambda^{ab}$. By using
$\widetilde{g_{ab}}\widetilde{g^{bc}}=\delta_{a}^{c}$, we obtain
\begin{equation}\label{relationgammalambda}
     \alpha^{2}\Omega^{2}\delta_{a}^{c}+ \alpha^{2}\gamma_{a}^{c}+\Omega^{2}\lambda_{a}^{c}+
     {\gamma}{_{ab}}\lambda^{bc}=\delta_{a}^{c}
\end{equation}
if the deformations are conformal transformations, we  have
$\alpha=\Omega^{-1}$, so  assuming such a condition, one obtain
the following matrix equation
\begin{equation}\label{relationgammalambda1}
    \alpha^{2}\gamma_{a}^{c}+\Omega^{2}\lambda_{a}^{c}+
     {\gamma}{_{ab}}\lambda^{bc}=0\,,
\end{equation}
and
 \begin{equation}\label{lambdaaa}
 (\delta_{a}^{b}+
     \Omega^{-2}{\gamma}_{a}^{b})\lambda_{b}^{c}=-\Omega^{-4}\gamma_{a}^{c}
 \end{equation}
 and finally
 \begin{equation}\label{lambdaaaa}
 \lambda_{b}^{c}=-\Omega^{-4}{{(\delta+
     \Omega^{-2}{\gamma})^{-1}}}{^{a}_{b}}\gamma_{a}^{c}
 \end{equation}
where   ${(\delta+ \Omega^{-2}{\gamma})^{-1}}$ is the inverse
tensor of $(\delta_{a}^{b}+
     \Omega^{-2}{\gamma}_{a}^{b})$.

To each matrix $\Phi^{A}_{\phantom{A}B}$,  we can associate a
(1,1)-tensor $\phi^{a}_{\phantom{a}b}$  defined by
\begin{equation}\label{3.1}
  \phi^{a}_{\phantom{a}b}= \Phi^{A}_{\phantom{A}B}\omega^{B}_{b}e_{A}^{a}
\end{equation}
such that
\begin{equation}\label{3.2}
     \widetilde{g}_{ab}=g_{cd}\phi^{c}_{\phantom{c}a}\phi^{d}_{\phantom{d}b}
\end{equation}
which can  be decomposed as in Eq.(\ref{split}). Vice-versa from a
(1,1)-tensor $\phi^{a}_{\phantom{a}b}$, we can define a matrix of
scalar fields as
\begin{equation}\label{3.3}
     \phi^{A}_{\phantom{A}B} =  \phi^{a}_{\phantom{a}b} \omega_{a}^{A}e_{B}^{b}.
\end{equation}

The  Levi Civita connection corresponding to the metric
(\ref{splits}) is related to the original connection by the
relation (see the Appendix for details)
\begin{equation}\label{1}
     {\widetilde{\Gamma}}{^{c}_{ab}}=  {\Gamma}{^{c}_{ab}} + {C}{^{c}_{ab}}
\end{equation}
(see \cite{wald}), where
\begin{equation}\label{2}
  {C}{^{c}_{ab}}=2\widetilde{g}^{cd}{g}{_{d(a}}{\nabla}{_{b)}} \Omega -g_{ab}\widetilde{g}^{cd} \nabla_{d}\Omega +\frac{1}{2} \widetilde{g}^{cd}\left( \nabla_{a}\gamma_{db}+\nabla_{b}\gamma_{ad}-\nabla_{d}\gamma_{ab}\right).
\end{equation}
Therefore, in a deformed space-time, the connection deformation
acts like a force that deviates  the test particles from the
geodesic motion in the unperturbed space-time. As a matter of fact
the geodesic equation for the deformed space-time

\begin{equation}\label{geodesics1}
     \frac{d^{\,2}x^{c}}{d\lambda^{2}}+ \tilde{\Gamma}^{c}_{\phantom{c}ab}\frac{dx^{a}}{d\lambda}
     \frac{dx^{b}}{d\lambda}=0
\end{equation}
becomes

\begin{equation}\label{geodesics2}
     \frac{d^{\,2}x^{c}}{d\lambda^{2}}+ \Gamma^{c}_{\phantom{c}ab}\frac{dx^{a}}{d\lambda}
     \frac{dx^{b}}{d\lambda}=-C^{c}_{\phantom{c}ab}\frac{dx^{a}}{d\lambda}
     \frac{dx^{b}}{d\lambda}.
\end{equation}

The  deformed Riemann curvature tensor is then
\begin{equation}\label{deformedcurvature}
    \widetilde{R}_{abc}^{\phantom{abc}d}=  R_{abc}^{\phantom{abc}d}+
    \nabla_{b} C ^{d}_{\phantom{d}ac}-\nabla_{a}C^{d}_{\phantom{d}bc}+
     C^{e}_{\phantom{e}ac} C^{d}_{\phantom{d}be}- C^{e}_{\phantom{e}bc}C^{d}_{\phantom{d}ae},
\end{equation}
while the deformed Ricci tensor obtained by contraction is
\begin{equation}\label{deformedricci}
  \widetilde{R}_{ab}= R_{ab} + \nabla_{d} C ^{d}_{\phantom{d}ab}-\nabla_{a}C^{d}_{\phantom{d}db}+ C^{e}_{\phantom{e}ab} C^{d}_{\phantom{d}de}- C^{e}_{\phantom{e}db}C^{d}_{\phantom{d}ae}
\end{equation}
and the curvature scalar
\begin{equation}\label{deformedcurvaturescalar}
  \widetilde{R}= \widetilde{g}^{ab} \widetilde{R}_{ab}=
  \widetilde{g}^{ab}{R}_{ab}+\widetilde{g}^{ab}
  \left[\nabla_{d} C ^{d}_{\phantom{d}ab}-\nabla_{a}C^{d}_{\phantom{d}db}+
  C^{e}_{\phantom{e}ab} C^{d}_{\phantom{d}de}- C^{e}_{\phantom{e}db}C^{d}_{\phantom{d}ae}\right]
\end{equation}

From  the above curvature quantities, we obtain finally the
equations for the deformations. In the vacuum case,  we simply
have
\begin{equation}\label{eeq}
   \widetilde{R}_{ab} = {R}_{ab}+\nabla_{d} C ^{d}_{\phantom{d}ab}-\nabla_{a}C^{d}_{\phantom{d}db}+ C^{e}_{\phantom{e}ab} C^{d}_{\phantom{d}de}- C^{e}_{\phantom{e}db}C^{d}_{\phantom{d}ae}=0
\end{equation}
where ${R}_{ab}$ must be regarded as a known function.  In
presence of matter, we consider the equation
\begin{equation}\label{eeqmatter}
 {R}_{ab}+\nabla_{d} C ^{d}_{\phantom{d}ab}-\nabla_{a}C^{d}_{\phantom{d}db}+ C^{e}_{\phantom{e}ab} C^{d}_{\phantom{d}de}- C^{e}_{\phantom{e}db}C^{d}_{\phantom{d}ae}= \widetilde{T}_{ab}-\frac{1}{2}\widetilde{g}_{ab}\widetilde{T}
\end{equation}
we are assuming, for the sake of simplicity $8\pi G=c=1$. This
last equation can be improved by considering the Einstein field
equations
\begin{equation}\label{undeformedeeq}
 {R}_{ab}=  T_{ab}-\frac{1}{2}g_{ab}T
\end{equation}
and then
\begin{equation}\label{definitivaeeqmatter}
     \nabla_{d} C ^{d}_{\phantom{d}ab}-\nabla_{a}C^{d}_{\phantom{d}db}+ C^{e}_{\phantom{e}ab} C^{d}_{\phantom{d}de}- C^{e}_{\phantom{e}db}C^{d}_{\phantom{d}ae}= \widetilde{T}_{ab}-\frac{1}{2}\widetilde{g}_{ab}\widetilde{T}-\left(T_{ab}-\frac{1}{2}g_{ab}T\right)
\end{equation}
 is the most general equation for  deformations.

\section{Metric deformations as perturbations and gravitational waves}

Metric deformations can be used to describe perturbations.  To
this aim we can simply consider the deformations

\begin{equation}\label{2defor}
     \Phi^{A}_{\phantom{A}B}=\delta^{A}_{\phantom{A}B}+\varphi^{A}_{\phantom{A}B}
\end{equation}
with
\begin{equation}\label{2piccolezza}
    |\,\varphi^{A}_{\phantom{A}B}|\ll 1,
\end{equation}
together with their derivatives
\begin{equation}\label{2piccolezza1}
    |\,\partial\varphi^{A}_{\phantom{A}B}|\ll 1\,.
\end{equation}
With this approximation, immediately we find the inverse relation
\begin{equation}\label{3defor}
     (\Phi^{-1})^{A}_{\phantom{A}B}\simeq\delta^{A}_{\phantom{A}B}-\varphi^{A}_{\phantom{A}B}.
\end{equation}
As a remarkable example,  we have that gravitational waves are
generally described, in linear approximation, as perturbations of
the Minkowski metric
\begin{equation}\label{linearapproximation}
     g_{ab}=\eta_{ab}+\gamma_{ab}.
\end{equation}
In our case, we can extend in a covariant way such an
approximation. If $ \varphi_{AB}$ is an antisymmetric matrix, we
have
\begin{equation}\label{covariantlinearapproximation}
\widetilde{g}_{ab}=g_{ab}+\gamma_{ab}
\end{equation}
where  the first order terms in $\varphi^{A}_{\phantom{A}B}$
vanish and $\gamma_{ab}$ is of second order
\begin{equation}\label{gammagw}
    \gamma_{ab}=\eta_{AB}\varphi^{A}_{\phantom{A}C}\varphi^{B}_{\phantom{B}D}\omega^{C}_{\phantom{C}a}
    \omega^{D}_{\phantom{D}b}.
\end{equation}
Consequently
\begin{equation}\label{controvariante}
    \widetilde{g}^{ab}=g^{ab}+\gamma^{ab}
\end{equation}
where
\begin{equation}\label{gam}
    \gamma^{ab}=\eta^{AB}(\varphi^{-1})_{\phantom{A}A}^{C}(\varphi^{-1})_{\phantom{B}B}^{D}e_{C}^{\phantom{A}a}
    e_{D}^{\phantom{B}b}.
\end{equation}
Let us consider the background metric $g_{ab}$, solution of the
Einstein equations in the vacuum
\begin{equation}\label{vacein}
    R_{ab}=0.
\end{equation}
We obtain the equation of perturbations considering  only the
linear terms in Eq.(\ref{eeq}) and neglecting the contributions of
quadratic terms. We find
\begin{equation}\label{approx1}
   \widetilde{R}_{ab} = \nabla_{d}  C ^{d}_{\phantom{d}ab}
   -\nabla_{a}C^{d}_{\phantom{d}db}=0\,,
\end{equation}
and, by the explicit form of $C ^{d}_{\phantom{d}ab}$, this
equation becomes
\begin{equation}\label{approx2}
    \left( \nabla_{d}\nabla_{a} {\gamma}{^{d}_{b}}+ \nabla_{d}\nabla_{b} {\gamma}{^{d}_{a}}-\nabla_{d}\nabla^{d} {\gamma}{_{ab}}\right)-
    \left( \nabla_{a}\nabla_{d} {\gamma}{^{d}_{b}}+ \nabla_{a}\nabla_{b} {\gamma}{^{d}_{d}}-
    \nabla_{a}\nabla^{d} {\gamma}{^{d}_{b}}\right)=0\,.
\end{equation}
Imposing the transverse traceless gauge on $\gamma_{ab}$ , i.e.
the standard gauge conditions
\begin{equation}\label{gauge1}
    \nabla^{a}\gamma_{ab}=0
\end{equation}
and
\begin{equation}\label{gauge2}
    \gamma={\gamma}{^{a}_{a}}=0
\end{equation}
Eq.(\ref{approx2}) reduces to
\begin{equation}\label{eqwald}
    \nabla_{b}\nabla^{b}\gamma_{ac}-2{R}{{^{\,b}_{ac}}^{d}}\gamma_{bd}=0\,,
\end{equation}
see also \cite{wald}. In our context, this equation is a
linearized equation for  deformations and it is straightforward to
consider perturbations and, in particular, gravitational waves, as
small deformations of the metric. This result can be immediately
translated into  the above scalar field matrix equations. Note
that such an equation can be applied to the conformal part of the
deformation, when the general decomposition is considered.

As an example, let us take into account the deformation matrix
equations applied to the Minkowski metric, when the deformation
matrix assumes the form (\ref{2defor}). In this case, the
equations (\ref{approx2}), become ordinary wave equations for
$\gamma_{ab}$. Considering the deformation matrices, these
equations become, for a tetrad field of constant vectors,
\begin{equation}\label{equationdeformation}
    \partial^{d} \partial_{d}\varphi^{C}_{\phantom{C}A}\varphi_{CB}+2\,
    \partial_{d}\varphi^{C}_{\phantom{A}A}  \partial^{d}\varphi_{CB}+ \varphi^{C}_{\phantom{C}A}\partial^{d}
    \partial_{d}\varphi_{CB}=0\,.
\end{equation}
The above gauge conditions are now
\begin{equation}\label{gauge1phi}
     \varphi_{AB}\varphi^{BA}=0
\end{equation}
and
\begin{equation}\label{gauge2phi}
     {e}{_{D}^{d}}\left[\partial_{d}\varphi_{CA}{\varphi}{^{C}_{B}}+
     \varphi_{CA}\partial_{d}{\varphi}{^{C}_{B}}\right]=0\,.
\end{equation}
This result shows that the gravitational waves can be fully
recovered starting from the scalar fields which describe the
deformations of the metric. In other words, such scalar fields can
assume the meaning of gravitational wave modes.

\section{Approximate Killing vectors}
Another important issue which can be addressed starting from
space-time deformations is related to the symmetries. In
particular, they assume a fundamental role in describing when a
symmetry is preserved or broken under the action of a given field.
In General Relativity, the Killing vectors are always related to
the presence of given space-time symmetries \cite{wald}.

Let us take an exact solution of the Einstein equations, which
satisfies the Killing equation
\begin{equation}\label{killing}
    ( L_{\mathbf{\xi}}g)_{ab}=0
\end{equation}
where $\mathbf{\xi}$, being the generator of an infinitesimal
coordinate transformation, is a Killing vector. If we take a
deformation of the metric with the scalar matrix
\begin{equation}\label{defor}
     \Phi^{A}_{\phantom{A}B}=\delta^{A}_{\phantom{A}B}+\varphi^{A}_{\phantom{A}B}
\end{equation}
with
\begin{equation}\label{piccolezza}
    |\,\varphi^{A}_{\phantom{A}B}|\ll 1\,,
\end{equation}
and
\begin{equation}\label{nokilling}
    ( L_{\mathbf{\xi}}\widetilde{g})_{ab}\neq 0\,,
\end{equation}
being

\begin{equation}\label{killingtetrad}
    ( L_{\mathbf{\xi}}e^{A})_{a}=0\,,
\end{equation}
we have
\begin{equation}\label{nokilling2}
    ( L_{\mathbf{\xi}} \varphi)^{A}_{\phantom{A}B} =
     \xi^{a}\partial_{a}\varphi^{A}_{\phantom{A}B}\neq 0\,.
\end{equation}
If  there is some region $\mathcal{D}$ of the deformed space-time
$\mathcal{M}_{deformed}$ where

\begin{equation}\label{nokilling3}
    |\,( L_{\mathbf{\xi}} \varphi)^{A}_{\phantom{A}B}|\ll 1
\end{equation}
we say that $\mathbf{\xi}$ is an {\it approximate Killing vector}
on $\mathcal{D}$. In other words, these approximate Killing
vectors allow to "control" the space-time symmetries under the
action of a given deformation.

\section{Discussion and conclusions}
In this paper, we have proposed  a novel definition of space-time
metric deformations parameterizing them in terms of scalar field
matrices. The main result is that deformations can be described as
extended conformal transformations. This fact gives a
straightforward physical interpretation of conformal
transformations: conformally related metrics can be seen as the
"background" and the "perturbed" metrics. In other words, the
relations between the Jordan frame and the Einstein frame can be
directly interpreted through the action of the deformation
matrices contributing to solve the issue of what the true physical
frame is \cite{faraoni,allemandi}.

Besides, space-time metric deformations can be immediately recast
in terms of perturbation theory allowing a completely covariant
approach to the problem of gravitational waves.

Results related to those presented here has been proposed in
\cite{Coll1,Coll:2001wy}. There it is shown that any metric in a
three dimensional manifold can be decomposed in the form
\begin{equation}\label{coll1}
     \widetilde{g}_{ab}= \sigma(x)h_{ab}+\epsilon s_{a}s_{b}
\end{equation}
where $h_{ab}$ is a metric with constant curvature, $\sigma(x)$ is
a scalar function,  $s_{a}$ is a three-vector and $\epsilon=\pm
1$. A relation has to be imposed between $\sigma$ and $s_{a}$ and
then  the metric can be defined, at most, by three independent
functions.

In a subsequent paper \cite{Llosa:2004uf}, Llosa and Soler showed
that (\ref{coll1}) can be generalized to arbitrary dimensions by
the form
\begin{equation}\label{coll2}
     \widetilde{g}_{ab}= \lambda(x)g_{ab}+\mu(x)F_{ac}g^{cd}F_{db}
\end{equation}
where $g_{ab}$ is a constant curvature metric, $F_{ab}$ is a
two-form, $\lambda(x)$ and $\mu(x)$ are two scalar functions.
These results are fully recovered and generalized from our
approach as soon as the deformation of a constant metric is
considered and suitable conditions on the tensor $\Theta_{AB}$ are
imposed.

In general, we have shown that, when we turn to the tensor
formalism, we can work with arbitrary metrics and arbitrary
deforming $\gamma_{ab}$ tensors. In principle, by arbitrary
deformation matrices, not necessarily real, we can pass from a
given metric to any other metric.   As an example, a noteworthy
result has been achieved by Newman and Janis \cite{Newman:1965tw}:
They showed that, through a complex coordinate transformation, it
is always possible to achieve a Kerr metric from a Schwarzschild
one. In our language, this means that a space-time deformation
allows to pass from a spherical symmetry to a cylindrical one.
Furthermore, it has been shown
\cite{Banados:1992wn,Banados:1992gq} that  three dimensional black
hole solutions can be found by identifying 3-dimensional anti-de
Sitter space on which acts a discrete subgroup of $SO(2,2)$.

In all these examples, the transformations which lead to the
results are considered as ``coordinate transformations''. We think
that this definition is a little bit misleading since one does not
covariantly perform the same transformations on {\it all} the
tensors defined on the manifold. On the other hand, our definition
of metric deformations and deformed manifolds can be
straightforwardly  related to the standard notion of perturbations
since, in principle, it works on a given region $\mathcal{D}$ of
the deformed space-time (see, for example,
\cite{Bardeen:1980kt,Mukhanov:1990me}).

\section{Appendix}
We can calculate the modified connection $\hat{\Gamma_{ab}^{c}}$
in many alternative ways. Let us introduce the tetrad $e_{A}$ and
cotetrad $\omega^{B} $ satisfying the orthogonality relation
\begin{equation}\label{ieomegaappendix}
   i_{e_{A}}\omega^{B}=\delta_{A}^{B}
\end{equation}
and the non-integrability condition (anholonomy)
\begin{equation}\label{1appendix}
d\omega^{A}=\frac{1}{2}\Omega_{BC}^{A}\omega^{B}\wedge\omega^{C}.
\end{equation}
The corresponding connection is
\begin{equation}\label{gammaappendix}
    \Gamma_{BC}^{A}=\frac{1}{2}\left(\Omega_{BC}^{A}-\eta^{AA'}\eta_{BB'}\Omega_{A'C}^{B'}-
    \eta^{AA'}\eta_{CC'}\Omega_{A'B }^{C'} \right)
\end{equation}
If we deform the metric as in (\ref{deformed}), we have two
alternative ways to write this expression: either writing the
``deformation'' of the metric in the space of tetrads or
``deforming''   the tetrad field as in the following expression
\begin{equation}\label{deformedagainappendix}
 \hat{g} =\eta_{AB}\Phi^{A}_{\phantom{A}C}\Phi^{B}_{\phantom{B}D}\,\omega^{C}
\omega^{D} =
  \gamma_{AB} \,\omega^{A}
\omega^{B}=\eta_{AB} \,\hat{\omega}^{A}\hat{ \omega}^{B}.
\end{equation}
In the first case,  the contribution of the Christoffel symbols,
constructed by the metric $\gamma_{AB}$, appears
$$ \hat{\Gamma}_{BC}^{A}=\frac{1}{2}\left(\Omega_{BC}^{A}-\gamma^{AA'}\gamma_{BB'}\Omega_{A'C}^{B'}-
    \gamma^{AA'}\gamma_{CC'}\Omega_{A'B }^{C'} \right) $$
\begin{equation}\label{hatgammaappendix}
+ \frac{1}{2}\gamma^{AA'}\left(i_{ e_{C}}  d\gamma_{BA'}-
i_{e_{B}}  d\gamma_{CA'}
     - i _{e_{A'}}  d\gamma_{BC}\right)
\end{equation}
In the second case,  using (\ref{1appendix}), we can define the
new anholonomy objects $\hat{C}_{BC}^{A}$.
\begin{equation}\label{2appendix}
d\hat{\omega}^{A}=\frac{1}{2}\hat{\Omega}_{BC}^{A}\hat{\omega}^{B}\wedge\hat{\omega}^{C}.
\end{equation}
After some  calculations, we have
\begin{equation}\label{newhatcappendix}
\hat{\Omega}_{BC}^{A}=\Phi^{A}_{\phantom{A}E}
{\Phi^{-1}}_{\phantom{D}B}^{D}
{\Phi^{-1}}_{\phantom{F}C}^{F}\,{\Omega}_{DF}^{E}+2\Phi^{A}_{\phantom{A}F}e_{G}^{a}
\left({\Phi^{-1}}_{\phantom{G}[B}^{G}\partial_{a}{\Phi^{-1}}_{\phantom{F}C]}^{F}\right)
\end{equation}
As we are assuming a constant metric in  tetradic space, the
deformed connection is
\begin{equation}\label{gammahat1appendix}
   \hat{ \Gamma}_{BC}^{A}=\frac{1}{2}\left(\hat{\Omega}_{BC}^{A}-\eta^{AA'}\eta_{BB'}\hat{\Omega}_{A'C}^{B'}-
    \eta^{AA'}\eta_{CC'}\hat{\Omega}_{A'B }^{C'} \right).
\end{equation}

Substituting (\ref{newhatcappendix}) in (\ref{gammahat1appendix}),
the final expression of $\hat{ \Gamma}_{BC}^{A}$, as a function of
$\Omega_{BC}^{A}$, $\Phi^{A}_{\phantom{A}B}$,
${\Phi^{-1}}_{\phantom{D}C}^{D}$ and $e_{G}^{a}$ is

\begin{equation}\label{gammahat2appendix}
 \hat{ \Gamma}_{ABC}=\Delta_{ABC}^{DEF}\left[\frac{1}{2}
 \eta_{DG}\,\Phi^{G}_{\phantom{G}G'}
{\Phi^{-1}}_{\phantom{E'}E}^{E'}
{\Phi^{-1}}_{\phantom{F'}F}^{F'}\,\Omega^{G'}_{E'F'} +
\eta_{DK}\Phi^{K}_{\phantom{D}H}e_{G}^{a}
 {\Phi^{-1}}_{\phantom{G}
[E}^{G}\partial_{|a|}{\Phi^{-1}}_{\phantom{H}F] }^{H}\right]
\end{equation}
where
\begin{equation}\label{delta1}
\Delta_{ABC}^{DEF}=\delta_{A}^{D}\delta_{C}^{E}\delta_{B}^{F}-\delta_{B}^{D}\delta_{C}^{E}\delta_{A}^{F}+
\delta_{C}^{D}\delta_{A}^{E}\delta_{B}^{F}.
\end{equation}

\end{document}